\documentclass[showpacs,preprintnumbers,amsmath,amssymb]{revtex4}

\usepackage{graphicx}
\usepackage{dcolumn}
\usepackage{bm}

\begin{document}

\bibliographystyle{phcjp}

\title{\bf Transport properties for liquid silicon-oxygen-iron mixtures at Earth's core conditions}

\author{Monica Pozzo$^1$}
\author{Chris Davies$^2$}
\author{David Gubbins$^{2,3}$}
\author{Dario Alf\`{e}$^1$}%
\email{d.alfe@ucl.ac.uk}

\affiliation{
  $^1$Department of Earth Sciences, Department of Physics and
  Astronomy, London Centre for Nanotechnology and Thomas Young Centre@UCL, University
  College London, Gower Street, London WC1E 6BT, United Kingdom }
\affiliation{ $^2$School of Earth and Environment, University of Leeds, Leeds LS2 9JT, UK}
\affiliation{ $^3$Institute of Geophysics and Planetary Physics, Scripps Institution of Oceanography, University of California at San Diego, 9500 Gilman Drive no. 0225, La Jolla, California 92093-0225, USA}

\date{\today}

\begin{abstract}
  We report on the thermal and electrical conductivities of two liquid
  silicon-oxygen-iron mixtures (Fe$_{0.82}$Si$_{0.10}$O$_{0.08}$ and
  Fe$_{0.79}$Si$_{0.08}$O$_{0.13}$), representative of the composition
  of the Earth's outer core at the relevant pressure-temperature
  conditions, obtained from density functional theory calculations
  with the Kubo-Greenwood formulation. We find thermal conductivities
  $k$ =100 (160) W m$^{-1}$ K$^{-1}$, and electrical conductivities
  $\sigma = 1.1 (1.3) \times 10^6 \Omega^{-1}$ m$^{-1}$ at the top
  (bottom) of the outer core. These new values are between 2 and 3
  times higher than previous estimates, and have profound implications
  for our understanding of the Earth's thermal history and the
  functioning of the Earth's magnetic field, including rapid cooling
  rate for the whole core or high level of radiogenic elements in the
  core.  We also show results for a number of structural and dynamic
  properties of the mixtures, including the partial radial
  distribution functions, mean square displacements, viscosities and
  speeds of sound.

\end{abstract}

\maketitle

\section{Introduction}

Transport properties of the Earth's core are of great importance to
understand the thermal and magnetic behaviour of our planet.  It is
widely believed that the Earth's core is a mixture of iron and light
impurities~\cite{Poirer94}. First principles calculations indicate
that the liquid outer core contains a significant fraction of oxygen
(between 8 or 13 \%), and silicon and/or sulphur (between 8 and 10
\%)~\cite{alfe07}. Oxygen is expelled into the liquid upon freezing of
the inner core at inner-core boundary (ICB), which helps to drive
convection responsible for the generation of the Earth's magnetic
field. Vigorous convection keeps the outer core in a well-mixed state,
with a temperature distribution which closely follows an adiabatic
profile.  Heat flows from the bottom to the top of the core by thermal
conduction and by convection. High thermal conductivity reduces the
heat available to drive convection (in the limit of infinite thermal
conductivity the whole core would become isothermal and would not
convect).
 
The electrical resistivity - and hence its inverse, the electrical
conductivity $\sigma$ - determines the magnitude of Ohmic losses of
the electric currents in the outer core that generate the magnetic
field, and also the magnetic diffusion time (that is, the time that it
would take for the magnetic field to decay in the absence of a generating 
mechanism). First experimental
measurements of the electrical conductivity of pure iron combining
high pressures and temperatures date back to the late sixties.  They
were performed on shock-compressed iron up from a few GPa to about 140
GPa and provided values for $\sigma$ in the range 1.0-1.5 $\times 10^6
\Omega^{-1}$ m$^{-1}$ (see Ref.~\cite{Keeler69} and refs. therein;
also Ref.~\cite{Keeler71}). These results were in line with earlier
estimates of $\sigma \simeq 1 \times 10^6 \Omega^{-1}$ m$^{-1}$ based
on geophysical arguments~\cite{Elsasser46}.  A few decades later,
Secco and Schloessin~\cite{Secco89} measured the electrical
conductivity of pure solid and liquid Fe at pressures up to few GPa
and then extrapolated their results to high pressures and temperatures
typical of the outer core. Noting the effects that the extrapolation
to higher parameter values had on the function of the density of
states, they derived values for $\sigma$ in the range 0.66-0.83
$\times 10^6 \Omega^{-1}$ m$^{-1}$, which agrees well with the results
previously obtained by Keeler and Mitchell~\cite{Keeler69}, but is
slightly smaller than the value measured by Keeler and
Royce~\cite{Keeler71}. More recently, the electrical conductivity of
pure iron has been measured in shock compression experiments at high
pressures and temperatures (up to 208 GPa and 5220 K) by Bi et
al.~\cite{Bi02}.  Its value ranges between 0.76 and 1.45 $\times 10^6
\Omega^{-1}$ m$^{-1}$ by reducing the pressure from about 200 to 100
GPa. These findings are in agreement with the result of about 1.0
$\times 10^6 \Omega^{-1}$ m$^{-1}$ measured by Gomi et
al.~\cite{Gomi10} in analogous experiments performed with a
diamond-anvil cell under high static pressure (up to 65 GPa) at room
temperature. They are also close to the value $\sigma = 0.8 \times
10^6 \Omega^{-1}$ m$^{-1}$ estimated by Stacey and
Anderson~\cite{Stacey01}, by assuming that, for a pure metal, the
conductivity is constant on its melting curve.  First-principles density
functional theory (DFT) calculations of solid iron at low temperature and 
under pressure have been presented by Sha and Cohen~\cite{Sha11}. 
Using the Bloch-Gr\"{u}neisen formula they extrapolated their results up to 
the high pressures and temperatures studied by Bi et al., finding a
slightly larger value in the range 1.1-1.8 $\times 10^6 \Omega^{-1}$
m$^{-1}$.  Recently we have calculated the electrical conductivities
of pure liquid iron at the conditions of the Earth's outer core using
DFT with the Kubo-Greenwood formulation (DFT-KG), 
obtaining values in the range 1.4-1.6 $\times 10^6 \Omega^{-1}$ m$^{-1}$.~\cite{pozzo12} 
An independent calculation was also performed by de Koker et al.~\cite{Koker12} showing similar
values. These values agree well with the experimental findings of the
older shock wave measurements of Keeler and Royce~\cite{Keeler71}, but
they are larger than the experimental values reported by Bi et
al. above 120 GPa~\cite{Bi02}.
 
The conductivities of the mixtures with typical core compositions were
estimated by Stacey and Anderson~\cite{Stacey01} to be about $0.5
\times 10^6 \Omega^{-1}$ m$^{-1}$ in the case of FeNiSi liquid alloys,
by extrapolating measured resistivity values of FeSi alloys from
Matassov~\cite{Matassov77} experimental data.  Our preliminary results
of FeSiO liquid alloys at Earth's core conditions indicated only a
reduction of about $30$\% of the conductivity of the mixture compared
to that of pure iron~\cite{pozzo12}, suggesting therefore values above
$1 \times 10^6 \Omega^{-1}$ m$^{-1}$.

Estimates for the thermal conductivity $k$ for liquid iron and liquid
iron mixtures ranged between 25 and 60 $m^{-1}$
K$^{-1}$.~\cite{Stacey01,Davies07,Stacey07,Konopkova11}  Our recent
DFT-KG calculations for pure iron at core conditions provided values
significantly higher, as expected from the relation between the
thermal and the electrical conductivity as encoded in the
Wiedemann-Franz law~\cite{wiedemann1853}, which we found to be closely
followed~\cite{pozzo12}.

The DFT-KG method has been applied to a wide range of problems,
including low pressure systems (C, Na, Al, Ar, Ga,
Pb)~\cite{Galli89,Holender95,Silvestrelli97,Tretiakov04,Recoules05,Knider07},
high pressure (H, H-He mixtures and
Al)~\cite{Morales09,Morales10,Lorenzen11,Vlcek12} and ultrahigh
pressure (water)~\cite{French10}.  Here we report detailed DFT-KG
calculations for $\sigma$ and $k$ of liquid iron mixtures,
Fe$_{0.82}$Si$_{0.10}$O$_{0.08}$ and
Fe$_{0.79}$Si$_{0.08}$O$_{0.13}$, representative of Earth's core
composition.  We find $\sigma = 1.1-1.3 \times 10^6 \Omega^{-1}$
m$^{-1}$ and $k= 100-160$ W $m^{-1}$ K$^{-1}$, the two extremes in the
ranges corresponding to the top and the bottom of the core,
respectively.  These values are in close agreement with extrapolations
obtained from recent diamond-anvil-cell experimental measurements of
the electrical resistivity by Hirose et al.~\cite{Hirose11} and by
Gomi et al.~\cite{Gomi11,Gomi12} (who estimated the thermal
conductivity from the resistivity using the Wiedemann-Franz law). We
find the Wiedemann-Franz law to be closely followed also for the iron
mixtures, although with a lower value for the Lorenz parameter.

These new values for the conductivities of the liquid outer core have
profound implications for our understanding of the Earth's thermal
history and the generation of the Earth's magnetic field.

For the typical Earth's core mixtures investigated here we also show
results for the partial radial distribution functions, mean square
displacements, viscosities and the speeds of sound.

In Sec.~\ref{sec:techniques} we describe the techniques used in the
calculations. The following section contains our results, starting
with the pressure-temperature profile in the core in
Sec.~\ref{sec:pt}, the structural properties of the mixtures in
Sec.~\ref{sec:structure}, ionic transport properties in
Sec.~\ref{sec:it} and electronic transport properties in
Sec.~\ref{sec:et}. Sec.~\ref{sec:implications} includes a discussion
of the implications of our results for the Earth, and
Sec.~\ref{sec:conclusions} contains the conclusions.

\section{Techniques}\label{sec:techniques}

First principles simulations were performed using the {\sc vasp}
code~\cite{kresse96}, with the projector augmented wave (PAW)
method~\cite{blochl94,kresse99} and the Perdew-Wang~\cite{pw91}
functional (PW91). The PAW potential for oxygen, silicon and iron have
the $2s^22p^4$, $3s^23p^2$ and $3p^64s^13d^7$ valence electronic
configurations respectively, and the core radii were 0.79, 0.8 and
1.16~\AA. To calculate the electrical conductivity we also tested two
additional iron PAW potentials, with $4s^13d^7$ and $3s^2sp^64s^13d^7$
valence configurations and 1.16 and 0.85~\AA~core radii
respectively. The potentials with more semi-core states included in
valence only give contributions to the optical conductivity at high
frequencies, but provide the same dc conductivity. For this reason,
conductivities have been calculated with the PAW potential having the
$4s^13d^7$ valence electronic configuration. Single particle orbitals
were expanded in plane-waves with a cutoff of 400 eV. Electronic
levels were occupied according to Fermi-Dirac statistics, with an
electronic temperature corresponding to the temperature of the
system. An efficient extrapolation of the charge density was used to
speed up the {\em ab initio} molecular dynamics
simulations~\cite{alfe99a}, which were performed by sampling the
Brillouin zone (BZ) with the $\Gamma$ point only. The temperature was
controlled with a Nos\'{e} thermostat~\cite{andersen80} and the time
step was set to 1 fs.  We ran simulations for typically 22 ps, from
which we discarded the first ps to allow for equilibration.

In the adiabatic approximation the electrical conductivity of a liquid
can be computed by generating a set of ionic configurations $\{R_I\}$
sampling the relevant pressure-temperature conditions, calculating the
conductivity on each of these configurations and taking the average.
For the conductivities we used the first 5 ps of the simulations to
extract $N=40$ configurations $\{R_I; I=1,N\}$ equally spaced in
time. These $N$ configurations were then used to compute the
electrical conductivity via the Kubo-Greenwood formula as implemented
in {\sc vasp} by Desjarlais~\cite{Desjarlais02}.  We performed
simulations at several thermodynamic states spanning the conditions of
the Earth's core, details for the simulations used to compute the
conductivities will be given in Sec.~\ref{sec:et}.

\begin{figure}
\includegraphics[width=4.0in]{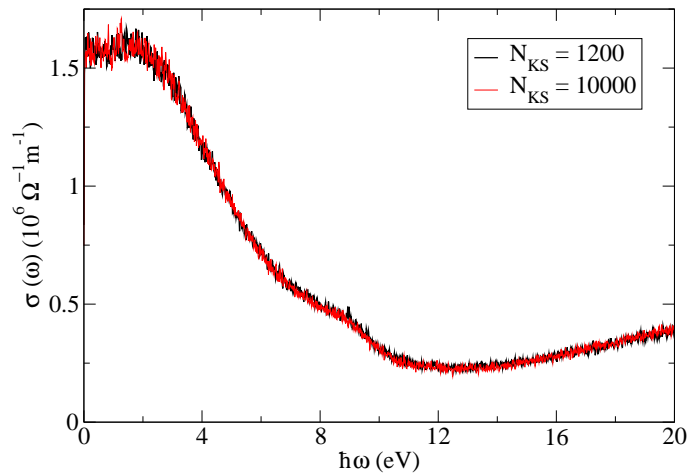}
\caption{\label{fig:sigmaomega} Optical conductivity $\sigma(\omega)$
  of liquid iron as function of energy computed using N$_{KS}=$10,000
  and N$_{KS}=$1,200 Kohn-Sham states for one configuration extracted
  from the ensemble at $p = 328$~GPa and $T = 6350$~K.}
\end{figure}

The KG formula for the frequency ($\omega$) dependent optical
conductivity for a system with $N_{I}$ ions at positions $R_I$ reads:
\begin{equation}\label{eqn:cond}
  \sigma_{\bf k}(\omega;R_I) = \frac{2\pi e^2 \hbar^2}{3m^2\omega\Omega}\sum_{i,j=1}^n \sum_{\alpha=1}^3 [F(\epsilon_{i,{\bf k}})-F(\epsilon_{j,{\bf k}})]|\langle \Psi_{j,{\bf k}} |\nabla_\alpha|\Psi_{i,{\bf k}}\rangle|^2\delta(\epsilon_{j,{\bf k}}-\epsilon_{i,{\bf k}}-\hbar\omega)
\end{equation}
where $e$ and $m$ are the electron charge and mass respectively,
$\hbar$ is the Plank's constant divided by $2\pi$, $\Omega$ is the
volume of the simulation cell and $n$ the number of Kohn-Sham
states. The $\alpha$ sum runs over the three spatial directions, which
in a liquid are all equivalent. $\Psi_{i,{\bf k}}$ is the Kohn-Sham
wavefunction corresponding to eigenvalue $\epsilon_{i,{\bf k}}$, and
$F(\epsilon_{i,{\bf k}})$ is the Fermi weight.  The $\delta$ function
is represented by a Gaussian, with a width chosen to be roughly equal
to the average spacing between the eigenvalues (0.01 eV for a 157-atom
system) weighted by the corresponding change in the Fermi
function~\cite{Desjarlais02}.  Integration over the Brillouin Zone
(BZ) is performed using standard methods~\cite{monkhorst76}, and the
frequency dependent conductivity is obtained by taking the thermal
average:
\begin{equation}
  \sigma(\omega) = \langle \sum_{\bf k}\sigma_{\bf k}(\omega;R_I)W({\bf k})\rangle
\end{equation}
where $W({\bf k})$ is the weighting factor for the point ${\bf k}$.
The dc conductivity $\sigma_0$ is given by the value of
$\sigma(\omega)$ in the limit $\omega \rightarrow 0$.
The optical conductivity must obey the sum rule:
\begin{equation}\label{eqn:sumrule}
S = \frac{2m\Omega}{\pi e^2 N_e}\int_0^\infty \sigma(\omega)d\omega = 1,
\end{equation}
where $N_e$ is the number of electrons in the simulation cell. As
reported earlier~\cite{pozzo12b}, to converge the sum rule for iron
one needs to include states up to $\sim 150$ eV above the Fermi
energy, which means using over 10,000 Kohn-Sham states for a typical
157 atoms simulation cell with the 8 valence electron PAW iron
potential. However, only states near the Fermi energy contribute to
the zero frequency limit of the optical conductivity. This is
illustrated in Fig.~\ref{fig:sigmaomega}, where we show
$\sigma(\omega)$ as function of $\omega$ computed using both 10,000
and 1,200 Kohn-Sham states for one configuration of liquid iron
extracted from the ensemble at $p = 328$~GPa and $T = 6350$~K. It is
clear that both calculations give the same dc conductivity, so we
decided to spot check the sum rule only in a limited number of cases,
and then use 1,200 Kohn-Sham states for the majority of the
calculations.

In a free electron liquid the electronic part of the thermal
conductivity $\kappa_0$ and the electrical conductivity $\sigma_0$ are
related by the Wiedemann-Franz law, $L= \kappa_0 / \sigma_0 T$, where
$L$ is the Lorenz number.  In a real liquid the validity of the
Wiedemann-Franz law is not necessarily expected, and in fact a number
of exceptions for metals at near ambient conditions are known (see
e.g. Kittel~\cite{kittel}).  Here we have directly calculated
$\kappa_0$ using the Chester-Thellung~\cite{chester61} (CS)
formulation of the Kubo-Greenwood formula, which reads:
\begin{equation}\label{eqn:t1}
  \kappa(\omega) = \frac{1}{e^2T}\left (L_{22}(\omega) - \frac{L_{12}(\omega)^2}{\sigma(\omega)} \right ),
\end{equation}
and $\kappa_0$ is the value of $\kappa(\omega)$ in the limit $\omega
\rightarrow 0$.  The kinetic coefficients $L_{lm}(\omega)$ are given
by~\cite{mazevet10}:
\begin{eqnarray}\label{eqn:t2}
  L_{lm}(\omega) = (-1)^{(l+m)}\frac{2\pi e^2 \hbar^2}{3m^2\omega\Omega}\sum_{i,j=1}^n \sum_{\alpha=1}^3 [F(\epsilon_{i,{\bf k}})-F(\epsilon_{j,{\bf k}})]|\langle \Psi_{j,{\bf k}} |\nabla_\alpha|\Psi_{i,{\bf k}}\rangle|^2 \times \\ \nonumber
  \times \left [\epsilon_{j,{\bf k}} - \mu\right ]^{(l-1)}\left [\epsilon_{i,{\bf k}} - \mu\right ]^{(m-1)}\delta(\epsilon_{j,{\bf k}}-\epsilon_{i,{\bf k}}-\hbar\omega),
\end{eqnarray}
where $\mu$ is the chemical potential. The implementation of the CS
formula in {\sc vasp} is also due to Desjarlais~\cite{Desjarlais02}.

We checked convergence of the conductivities with respect to the size
of the system by performing calculations with cubic simulation cells
including 67, 157 and 288 atoms and using up to 6 {\bf k}-points to
sample the BZ. We found that even the smallest cell sampled with the
single {\bf k}-point (1/4,1/4,1/4) gives results converged to better
than 1\%. We then decided to use 157-atom simulation cells.

\section{Results}
\subsection{Pressure-temperature profile}\label{sec:pt}
\begin{figure}
\includegraphics[width=4.0in]{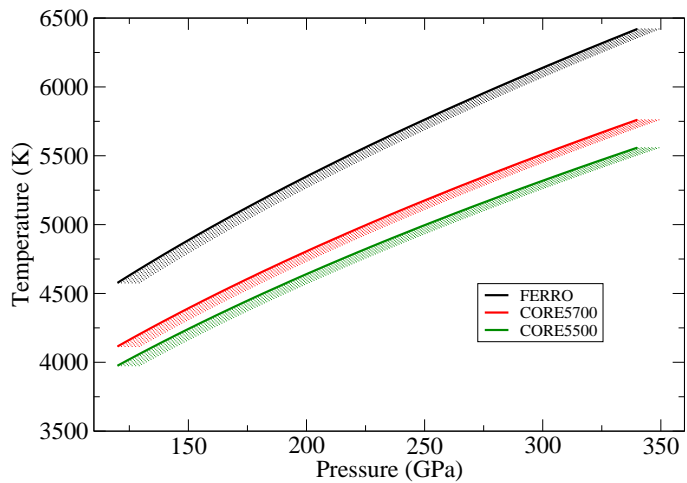}
\caption{\label{fig:tp} Adiabatic temperature profiles as function of
  pressure anchored by three possible ICB temperatures $T_{\rm
    ICB}$. Black curve (FERRO) corresponds to $T_{\rm ICB}= 6350$~K. Red
  curve (CORE5700) to $T_{\rm ICB}= 5700$~K, and green curve to
  $T_{\rm ICB}= 5500$~K. Low pressure edges of the bands (solid line)
  correspond to the raw DFT-PW91 pressures, high pressure edges to
  pressure corrected by 10 GPa, which is the typical DFT-PW91
  underestimate of the pressure for iron at core conditions (see text
  for details). }
\end{figure}

The Earth's outer core is convecting, and therefore it is assumed to
be in a well mixed state with an adiabatic pressure-temperature
($p,T$) profile. This can be determined by fixing the temperature $T$
at the ICB pressure $ p = 329$~GPa and following the line of constant
entropy up to the CMB pressure $p =136$~ GPa.  In Fig.~\ref{fig:tp} we
show the $p,T$ profiles that we used in the present work, obtained by
fixing three possible ICB temperatures: the melting temperature of
pure iron T = 6350 K~\cite{alfe02,alfe09} (FERRO), the melting
temperature of the mixture Fe$_{0.82}$Si$_{0.10}$O$_{0.08}$ T = 5700 K
(CORE5700)~\cite{alfe02b}, and the melting temperature of the mixture
Fe$_{0.79}$Si$_{0.08}$O$_{0.13}$ (CORE5500)~\cite{alfe07}. The
mixtures are two possible estimates for the composition of the Earth's
outer core, which match the {\it Preliminary Reference Earth Model}
(PREM) ICB density jump of 4.5\%~\cite{dziewonski81}, and the more
recent ICB density jump of 6.3\% proposed by Masters and
Gubbins~\cite{masters03}, respectively.  The adiabats are shown as
bands, with the low pressure (solid) edge corresponding to the actual
DFT-PW91 values, and the high pressure edge to pressures increased by
10 GPa, which is the approximate amount by which DFT-PW91
underestimates the pressure of solid iron at Earth's core
conditions~\cite{alfe02}.

\subsection{Structure}\label{sec:structure}
\begin{figure}
\includegraphics[width=4.0in]{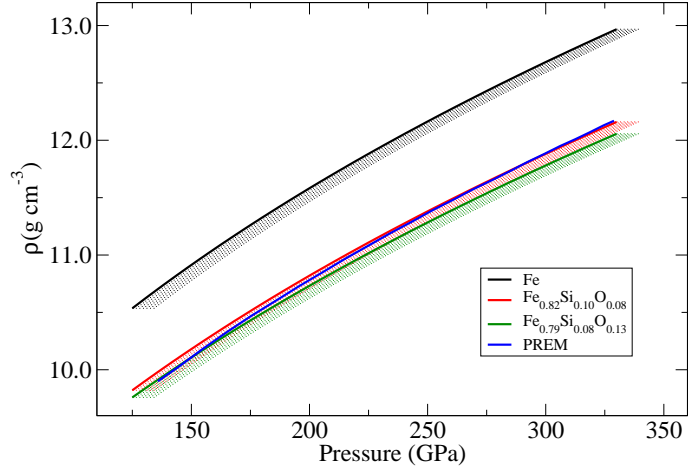}
\caption{\label{fig:rhop} Density profiles for pure iron (black), the
  Fe$_{0.82}$Si$_{0.10}$O$_{0.08}$ mixture (red), and the
  Fe$_{0.79}$Si$_{0.08}$O$_{0.13}$ mixture (green) on the respective FERRO,
  CORE5700 and CORE5500 pressure-temperature profiles displayed in
  Fig.~\ref{fig:tp}. Bands have the same meaning as in
  Fig.~\ref{fig:tp}. PREM density profile~\protect\cite{dziewonski81}
  is shown with blue line.}
\end{figure}

In Fig.~\ref{fig:rhop} we show the densities $\rho$ of the three
possible cases mentioned in Sec.~\ref{sec:pt} on the respective
adiabats. They are shown as bands also in this case, with the same
meaning as in Fig.~\ref{fig:tp}. The raw DFT-PW91 density of the
Fe$_{0.82}$Si$_{0.10}$O$_{0.08}$ (approximated by a simulation cell
containing 129 iron atoms, 16 silicon atoms and 12 oxygen atoms)
mixture matches the PREM density of the liquid side of the ICB by
construction~\cite{alfe02b}, while the mixture
Fe$_{0.79}$Si$_{0.08}$O$_{0.13}$ (125 irons, 12 silicons and 20
oxygens) has a slightly lower ICB density but appears to match quite
well the CMB density.

\begin{figure}
\includegraphics[width=4.0in]{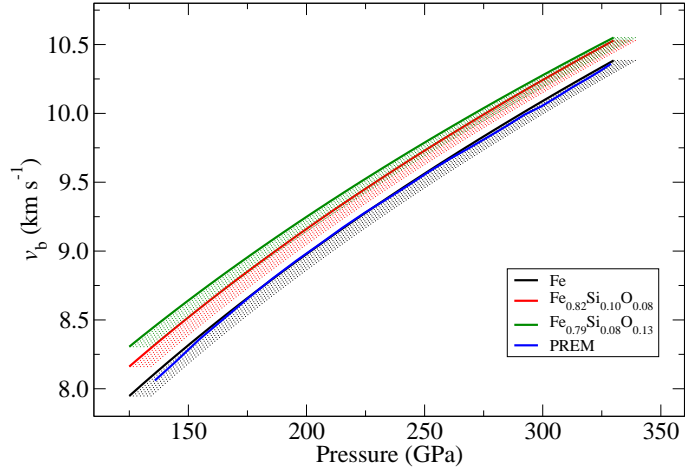}
\caption{\label{fig:vp} Bulk sound velocity for pure iron (black), the
  Fe$_{0.82}$Si$_{0.10}$O$_{0.08}$ mixture (red), and the
  Fe$_{0.79}$Si$_{0.08}$O$_{0.13}$ mixture (green) on the respective FERRO,
  CORE5700 and CORE5500 pressure-temperature profiles displayed in
  Fig.~\ref{fig:tp}. Bands have the same meaning as in
  Fig.~\ref{fig:tp}. The PREM bulk sound
  velocities~\protect\cite{dziewonski81} are shown in blue.}
\end{figure}

In Fig.~\ref{fig:vp} we plot the bulk sound velocities as function of
pressure for pure iron and for the two mixtures, compared with
PREM. These are defined as $v_b = \sqrt{(K_S/\rho)}$, where $K_S = -V
(dp/dV)_S$ is the isentropic bulk modulus, with $V$ and $S$ the volume
and the entropy of the system, respectively. To compute $K_S$ we
fitted the pressures computed along the adiabats to a Murnaghan
equation of state~\cite{murnaghan}:
\begin{equation}
p(V) = \frac{K_0}{K'_0}\left[\left(\frac{V_0}{V}\right)^{K'_0}-1\right]
\end{equation}
where $K_0$ is the zero pressure bulk modulus and $K'_0 =
(dK/dp)_{p=0}$.  Interestingly, the bulk sound velocities of pure
DFT-PW91 iron are very close to PREM, although the agreement is
worsened when the pressure correction of 10 GPa is applied to the
DFT-PW91 calculations. By contrast, the pressure corrected values for
the Fe$_{0.82}$Si$_{0.10}$O$_{0.08}$ mixture sit very close to
PREM. As expected, the combination of lower temperatures and densities
has the effect of increasing the bulk sound velocities.

\begin{figure}
\includegraphics[width=6.0in]{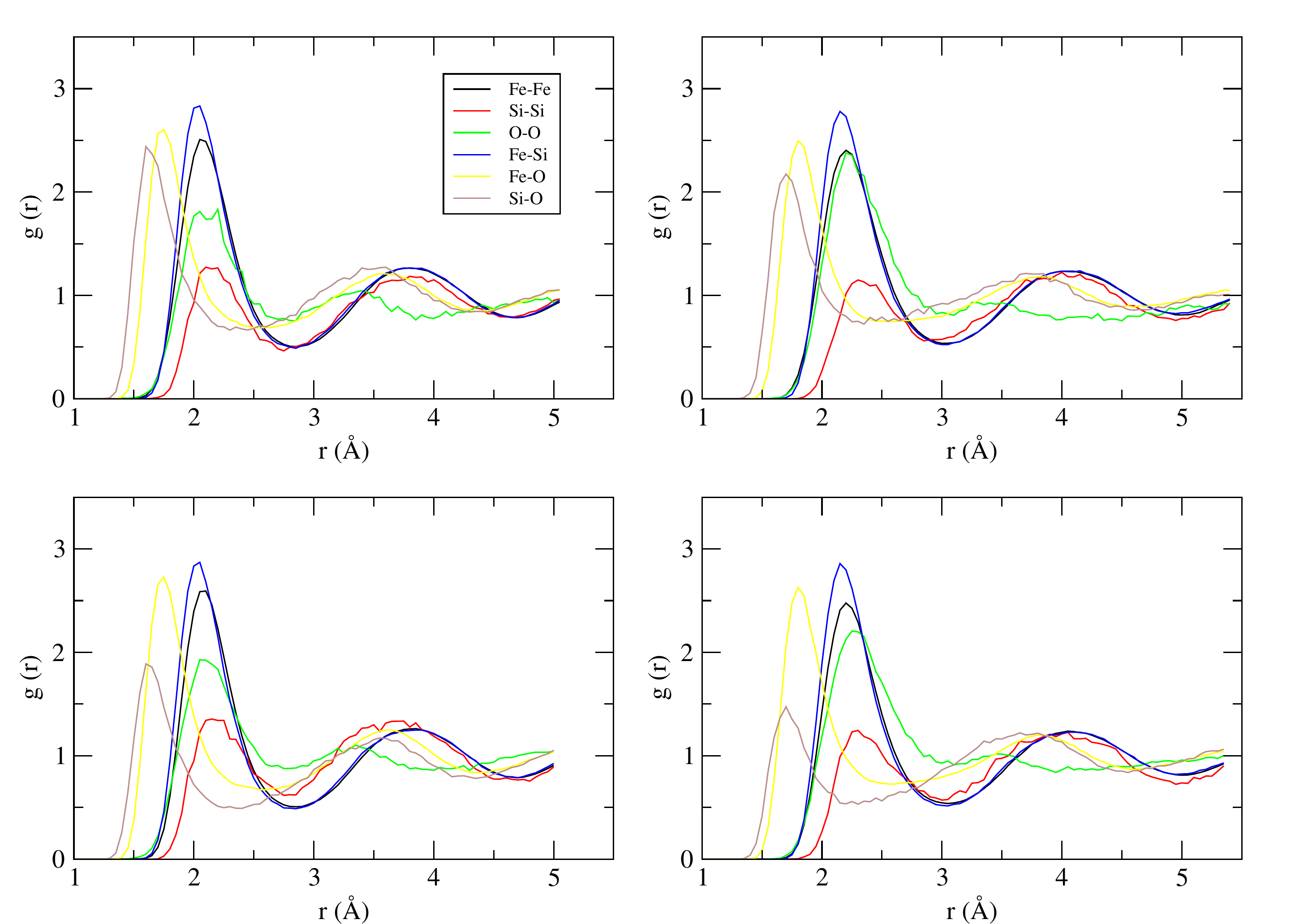}
\caption{\label{fig:gr} Partial radial distribution functions for the
  mixture Fe$_{0.82}$Si$_{0.10}$O$_{0.08}$ close to ICB (a) and CMB
  (b) conditions, and those for the mixture
  Fe$_{0.79}$Si$_{0.08}$O$_{0.13}$ close to ICB (c) and CMB (d)
  conditions (see text for details).}
\end{figure}

In Fig.~\ref{fig:gr} we show the partial radial distribution functions
(rdf) $g_{FeFe}(r)$, $g_{SiSi}(r)$, $g_{OO}(r)$, $g_{FeSi}(r)$,
$g_{FeO}(r)$ and $g_{SiO}(r)$. They are defined so that, by sitting on
an atom of the species $\alpha$, the probability of finding an atom of
the species $\beta$ in the spherical shell $(r, r+dr)$ is $4\pi
r^2n_\beta g_{\alpha\beta}(r)dr$, where $n_\beta$ is the number
density of the species $\beta$. Fig.~\ref{fig:gr} (a)
shows the partial rdfs for the mixture
Fe$_{0.82}$Si$_{0.10}$O$_{0.08}$ at $(p,T)=(329~{\rm GPa}, 5700~ {\rm
  K})$. In agreement with our previous simulations on FeO
mixtures~\cite{alfe99}, the present partial rdfs show that the
distance between neighbouring iron and oxygen atoms (obtained from the
position of the first peak of $g_{FeO}(r)$ at $\sim 1.7$~\AA) is
significantly shorter than the iron-iron distance ($\sim 2.1$~\AA) or
the oxygen-oxygen distance ($\sim 2.1$~\AA) . This indicates that
oxygen atoms have two effective radii, one for the interaction with
themselves, and a different one for the interaction with an iron
atom. The relatively large oxygen-oxygen distance suggests than oxygen
atoms effectively repel each other at the typical single and double
bond oxygen distances (1.47 \AA~ and 1.21 \AA,
respectively)~\cite{alfe99}. The situation is rather different for the
silicon-silicon and the iron-silicon distances, as the first peaks of
$g_{SiSi}(r)$ and $g_{FeSi}(r)$ are roughly in the same place, showing
that iron and silicon atoms have one single effective radius when
interacting with each other or with themselves, and also that this
effective radius is similar for the two atoms.  For the silicon-oxygen
interaction, the position of the first peak of $g_{SiO}(r)$ at $\sim
1.6$~\AA~is at slightly shorter distance than that of $g_{FeO}(r)$,
indicating that the silicon-oxygen bond is shorter and probably
stronger when compared to the iron-oxygen bond. Although it could
be suggested that oxygen and silicon in liquid iron may precipitate
out as SiO$_2$, the simulations provided no evidence of any phase
separation or departure from a well mixed liquid. In Fig.~\ref{fig:gr}
(b) we show the partial rdfs for the same mixture at $(p,T)=(134~{\rm
  GPa}, 4260~ {\rm K})$ (close to CMB conditions), and in
Fig.~\ref{fig:gr} (c,d) those for the mixture
Fe$_{0.79}$Si$_{0.08}$O$_{0.13}$ at $(p,T)=(328~{\rm GPa}, 5500~ {\rm
  K})$ and $(p,T)=(134~{\rm GPa}, 4112~ {\rm K}$), respectively.
There are no significant differences between the corresponding partial
rdfs for the two mixtures, and the only effect of moving from CMB to
ICB conditions is that of decreasing the height of the $g_{OO}(r)$
peak and increasing the height of the $g_{SiO}(r)$ peak, which implies
that the oxygen-silicon coordination number is slightly increased with
pressure.  No noticeable pressure effect is observed for the
iron-oxygen interactions.

\subsection{Ionic transport}\label{sec:it}

\begin{figure}
\includegraphics[width=4.0in]{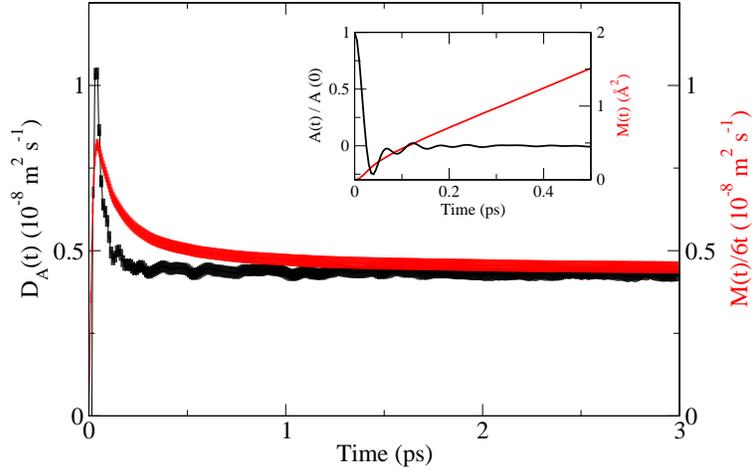}
\caption{\label{fig:m} Self-diffusion coefficient as a function of
  time obtained from the time dependent mean square displacement
  $M(t)$ (red) and the velocity-velocity autocorrelation function
  $D_A(t)$(black) (see text details). Also shown in the inset the
  value of $M(t)$ and $A(t)/A(0)$ as a function of time. Results
  correspond to the iron atoms in the Fe$_{0.82}$Si$_{0.10}$O$_{0.08}$
  mixture at ICB conditions.}
\end{figure}

In this section we describe the ionic dynamical properties of the
system, including atomic self-diffusion coefficients and viscosities.
The self-diffusion coefficient $D$ can be obtained from the asymptotic
slope of the time dependent mean square displacement (MSD) $M(t)$ in
the long time limit $t \rightarrow \infty$:
\begin{eqnarray}
  M(t) = \frac{1}{N_I}\sum_{i=1}^{N_I} \langle|{\bf r_i}(t+t_0) - {\bf r_i}(t_0)|^2\rangle \nonumber \\
  D = \frac{1}{6}\lim_{t\rightarrow \infty} \frac{dM(t)}{dt} \nonumber \\
\end{eqnarray}
where $\langle \cdot \rangle $ has the meaning of thermal average,
which is computed as time average over different origins $t_0$ along
the molecular dynamics simulation, ${\bf r_i}(t)$ is the vector position at time t
of the i-th atom, and $N_I$ is the number of ions. In the inset of Fig.~\ref{fig:m} we
show the value of $M(t)$ as function of time for the iron atoms in the
Fe$_{0.82}$Si$_{0.10}$O$_{0.08}$ mixture at ICB conditions. It is
clear that after a transient of $\sim 0.2$~ps the linear behaviour of
$M(t)$ is well established. The initial part of the transient is due
to atoms moving freely before collisions begin to occur, and for this
reason the MSD increases as the square of time.

\begin{figure}
\includegraphics[width=4.0in]{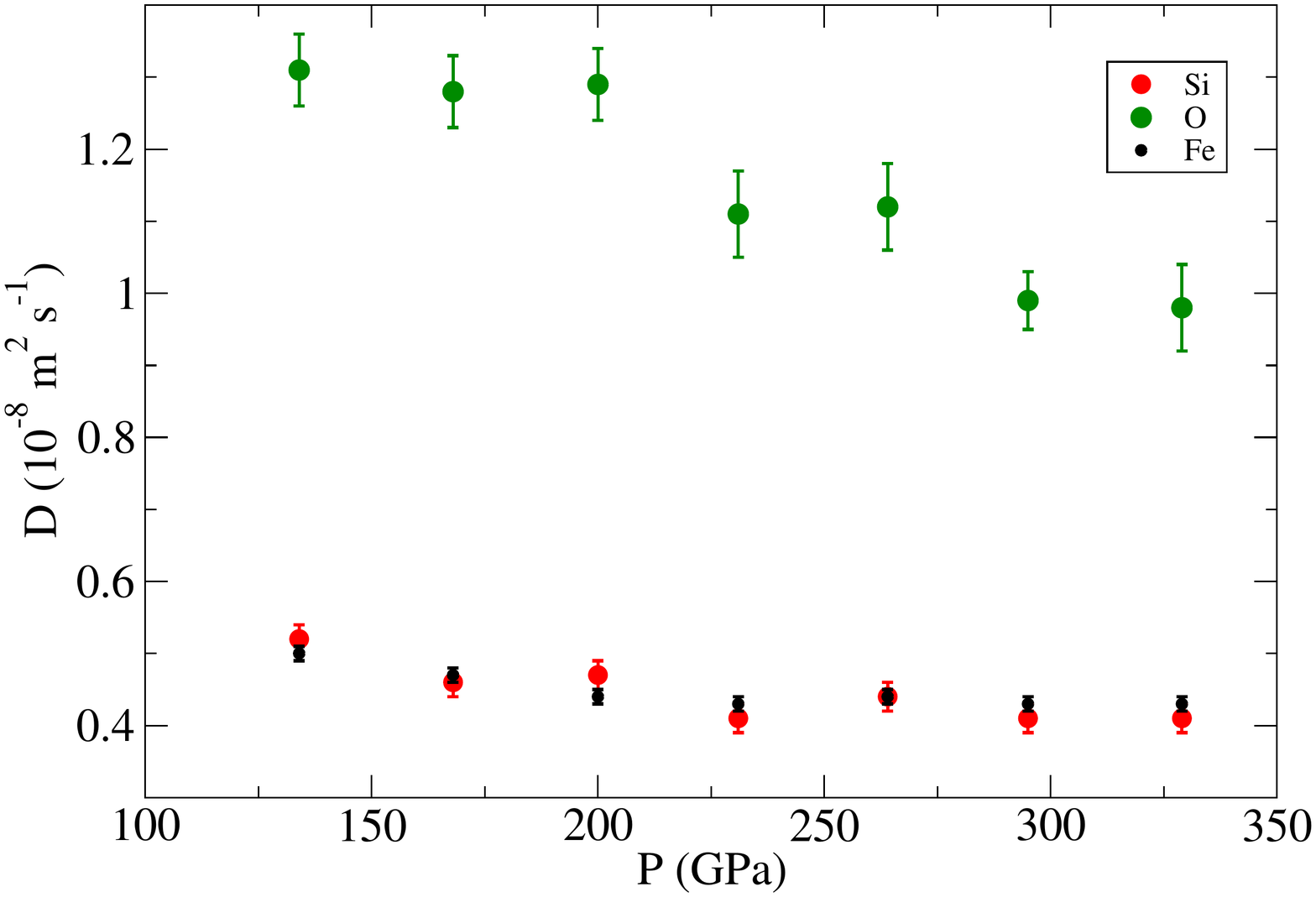}
\caption{\label{fig:D5700} Self diffusion coefficients of Fe, Si and O
  atoms vs pressure for the Fe$_{0.82}$Si$_{0.10}$O$_{0.08}$
  mixture. Temperatures are determined from the CORE5700 adiabat at
  the corresponding pressures.}
\end{figure}

\begin{figure}
\includegraphics[width=4.0in]{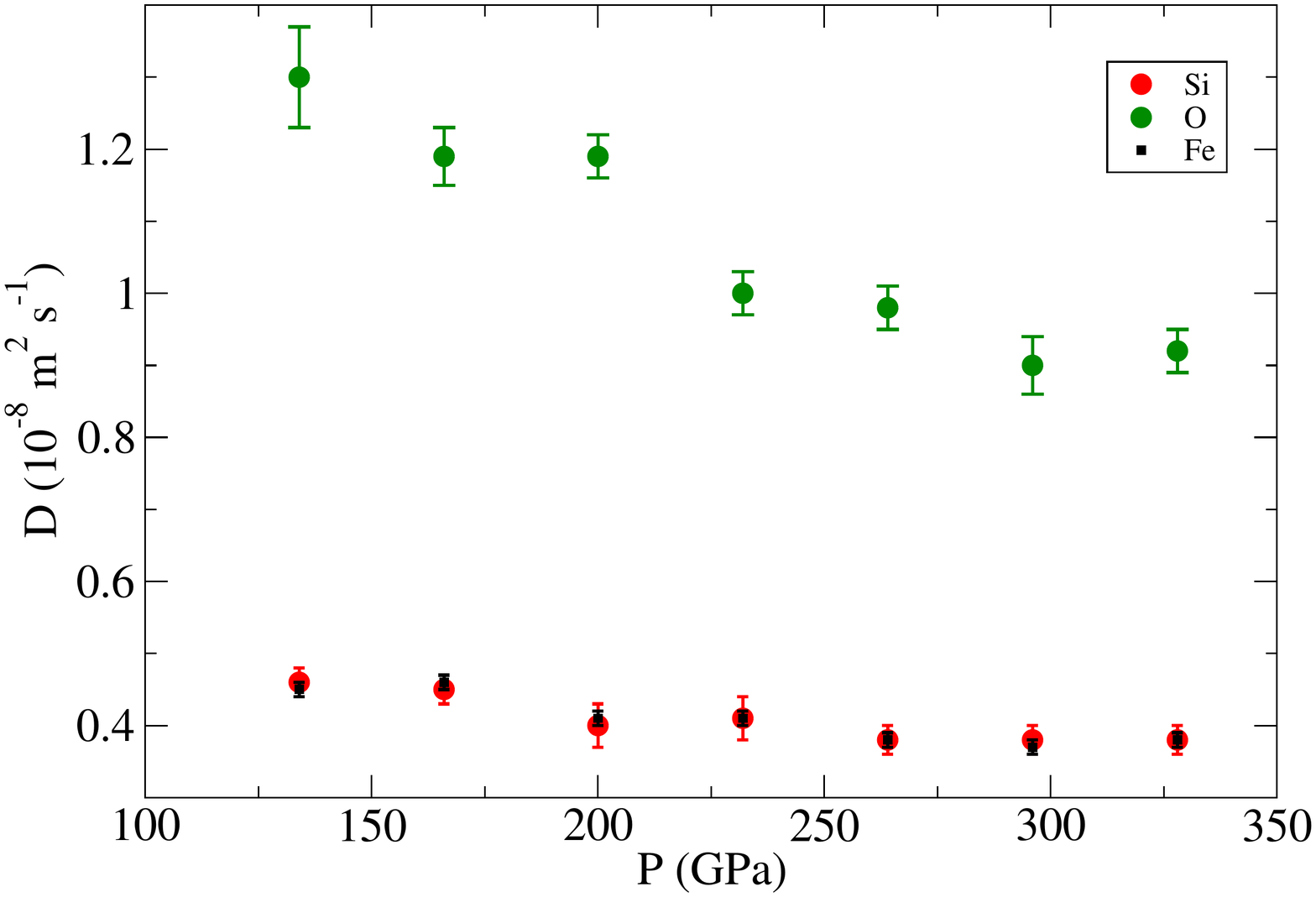}
\caption{\label{fig:D5500} Self diffusion coefficients of Fe, Si and O
  atoms vs pressure for the Fe$_{0.79}$Si$_{0.08}$O$_{0.13}$
  mixture. Temperatures are determined from the CORE5500 adiabat at
  the corresponding pressures.}
\end{figure}

An alternative route to the self-diffusion coefficient is through the
Green-Kubo (GK) relations, which relate transport coefficients and
correlation functions~\cite{allen}. The self-diffusion coefficient $D$
is given by the integral of the velocity-velocity autocorrelation
function (VACF) $A(t)$:
\begin{eqnarray}
  A(t) = \langle {\bf v}_i(t+t_0) \cdot {\bf v}_i(t_0)\rangle \nonumber \\
  D_A(t) = \frac{1}{3}\int_0^t d\tau A(\tau) \nonumber \\ 
  D = \lim_{t\rightarrow \infty} D_A (t) \nonumber \\
\end{eqnarray}
The function $A(t)/A(0)$ is also plotted in the inset of
Fig.~\ref{fig:m}, where it can be seen that correlations quickly decay
to zero as soon as the atoms begin to collide with each other.
In the main part of Fig.~\ref{fig:m} we show the value of $M(t)$
divided by $6t$, and that of $D_A(t)$ as function of time, together
with their statistical errors computed by analysing the scattering of
the MSD and VACF of each atom. The two methods provide the same
value for $D$ in the limit of large $t$, with $D_A(t)$ converging
faster. In Fig.~\ref{fig:D5700} and ~\ref{fig:D5500} we show the values
of the self diffusion coefficients of iron, silicon and oxygen for the
two mixtures. The values are almost constant along the adiabats, with
diffusion slowing down only marginally with pressure. Interestingly,
iron and silicon have very similar diffusion coefficients, despite a
factor of two difference in their masses, while oxygen has a diffusion
coefficient which is more than two times bigger. No significant
differences are observed for the two different mixtures.

\begin{figure}
\includegraphics[width=4.0in]{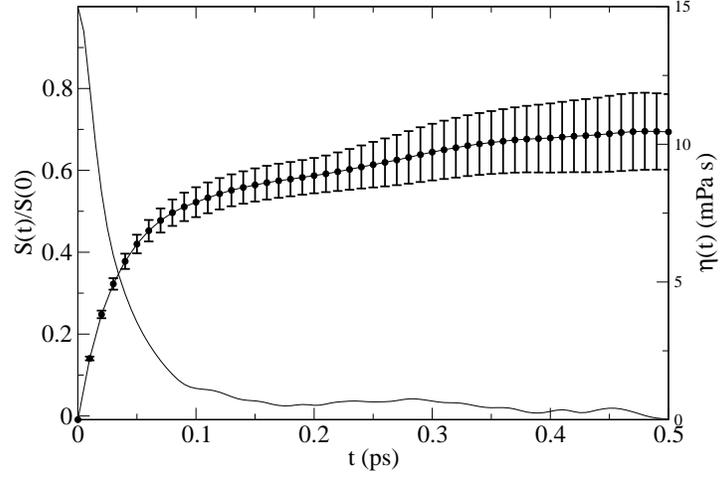}
\caption{\label{fig:visct} Average over the five independent
  components of the autocorrelation function of the traceless stress
  tensor $S(t)$, normalised by dividing by $S(0)$. Also shown is the
  viscosity integral $\eta(t)$ with its statistical error. Results
  correspond to the Fe$_{0.82}$Si$_{0.10}$O$_{0.08}$ mixture at ICB
  conditions (see text for details).}
\end{figure}

\begin{figure}
\includegraphics[width=4.0in]{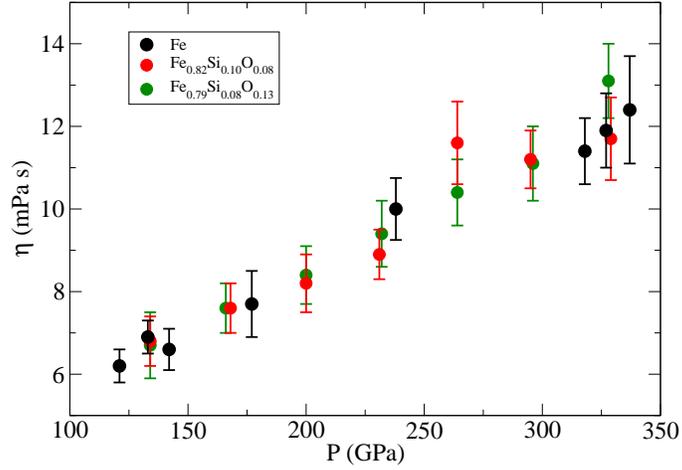}
\caption{\label{fig:viscp} Viscosity as a function of pressure for the
  pure iron on the FERRO adiabat, and for the mixtures
  Fe$_{0.82}$Si$_{0.10}$O$_{0.08}$ and
  Fe$_{0.79}$Si$_{0.08}$O$_{0.13}$ on the CORE5700 and CORE5500
  adiabats, respectively. }
\end{figure}

To compute the viscosities of the mixtures we used the Green-Kubo
relation, which relates the shear viscosity $\eta$ to the integral of
the autocorrelation function of the off-diagonal components of the
stress tensor (SACF) $C_{\alpha\beta}(t)= \langle
P_{\alpha\beta}(t+t_0)P_{\alpha\beta}(t_0)\rangle$, where
$P_{\alpha\beta}$ is an off-diagonal component of the stress tensor,
with $\alpha$ and $\beta$ indicating Cartesian components. There are
five independent components of the traceless stress tensor: $P_{xy}$,
$P_{yz}$, $P_{zx}$, $\frac{1}{2}(P_{xx}-P_{yy})$ and
$\frac{1}{2}(P_{yy}-P_{zz})$~\cite{alfe98}. In a liquid these five
components are equivalent, and they can all be used to improve the
statistical accuracy of the viscosity integral. The shear viscosity
$\eta$ is then obtained from:
\begin{eqnarray} 
  \eta(t) = \frac{V}{k_{B}T}\int_0^t d\tau S(\tau) \nonumber \\
  \eta = \lim_{t\rightarrow \infty} \eta(t) \nonumber \\ 
\end{eqnarray}
where $k_B$ is the Boltzmann constant and
$S(t)=\frac{1}{5}[C_{xy}(t)+C_{yz}(t)+C_{zx}+
\frac{1}{2}(C_{xx}(t)-C_{yy}(t))+\frac{1}{2}(C_{yy}(t)-C_{zz}(t))]$.
In Fig.~\ref{fig:visct} we plot $S(t)/S(0)$ and $\eta(t)$ for the
Fe$_{0.82}$Si$_{0.10}$O$_{0.08}$ mixture at ICB conditions. To obtain
the shear viscosity $\eta$ we need to integrate $S(t)$ to $t=\infty$,
however, it is clear that the $S(t)$ has decayed to zero after $\sim
0.2$~ps, after which one only integrates statistical noise. For this
reason, we decided to stop the integration at $t=0.2$~ps. The
viscosity integral $\eta(t)$ is also plotted in Fig.~\ref{fig:visct},
together with its error bar estimated by the scattering of the five
independent components of the traceless SACF. To improve the
statistics on the estimate of the statistical error on $\eta(t)$ we
split the simulations in two independent chunks.  The viscosities are
plotted in Fig.~\ref{fig:viscp} for the two mixtures
on the corresponding CORE5700 and CORE5500 adiabats. Their values
range between $\sim 7$ mPa~s at the CMB and $\sim 12$ mPa~s at the
ICB, with little difference between the viscosities of the two
mixtures. We also report the viscosities of pure iron on the FERRO
adiabat, which are consistent with those reported
before~\cite{dewijs98,alfe00}, and they have roughly the same values
as those of the mixtures.

To conclude this section we also report the ionic component of the
thermal conductivity of pure iron, $\kappa_{ion}$, computed using a
simple pair potential, which describes the energetics and the
structural and dynamical properties of iron very
accurately~\cite{alfe00}. The pair potential has the form $U({\bf
  r}_1, \dots, {\bf r}_{N_I}) = 4\sum_{i=1, i<j}^{N_I} (\sigma/|{\bf
  r}_i - {\bf r}_j|)^\alpha$, where $\{{\bf r}_i\}$ are the Cartesian
coordinates of the atoms, $\alpha = 5.86$, and $\sigma = 1.77$~\AA.
The Kubo-Green formula for the ionic thermal conductivity reads:
\begin{equation}
  \kappa_{ion} = \frac{1}{3\Omega k_BT^2}\int_0^\infty dt \langle {\bf j}(t+t_0)\cdot{\bf j}(t_0) \rangle
\end{equation}
where the microscopic heat current is given by:
\begin{equation} 
{\bf j}(t) = \sum_{i=1}^{N_I}{\bf v}_i\epsilon_i +
  \frac{1}{2}\sum_{i,j=1; j\ne j}^{N_I}{\bf r}_{ij}({\bf F}_{ij}\cdot
  {\bf v}_i)
\end{equation}
with ${\bf v}_i$ the velocity of atom $i$, ${\bf r}_{ij}$ the vector
distance between atom $i$ and atom $j$, and ${\bf F}_{ij}$ the force
on atom $i$ due to atom $j$ from the pair potential. The on-site
energy $\epsilon_i$ is:
\begin{equation} 
  \epsilon_i = \frac{1}{2}m|{\bf v}_i|^2 + \frac{1}{2}\sum_{j=1}^{N_I}4\left(\frac{\sigma}{|{\bf
        r}_i - {\bf r}_j|}\right )^\alpha
\end{equation}
with $m$ being the mass of the atoms. The calculated values range
between 2.5 and 4 W m$^{-1}$ K$^{-1}$, which compared to the
electronic contribution to the thermal conductivity (see next section)
are completely negligible.

\subsection{Electronic transport}\label{sec:et}

\begin{figure}
\includegraphics[width=4.0in]{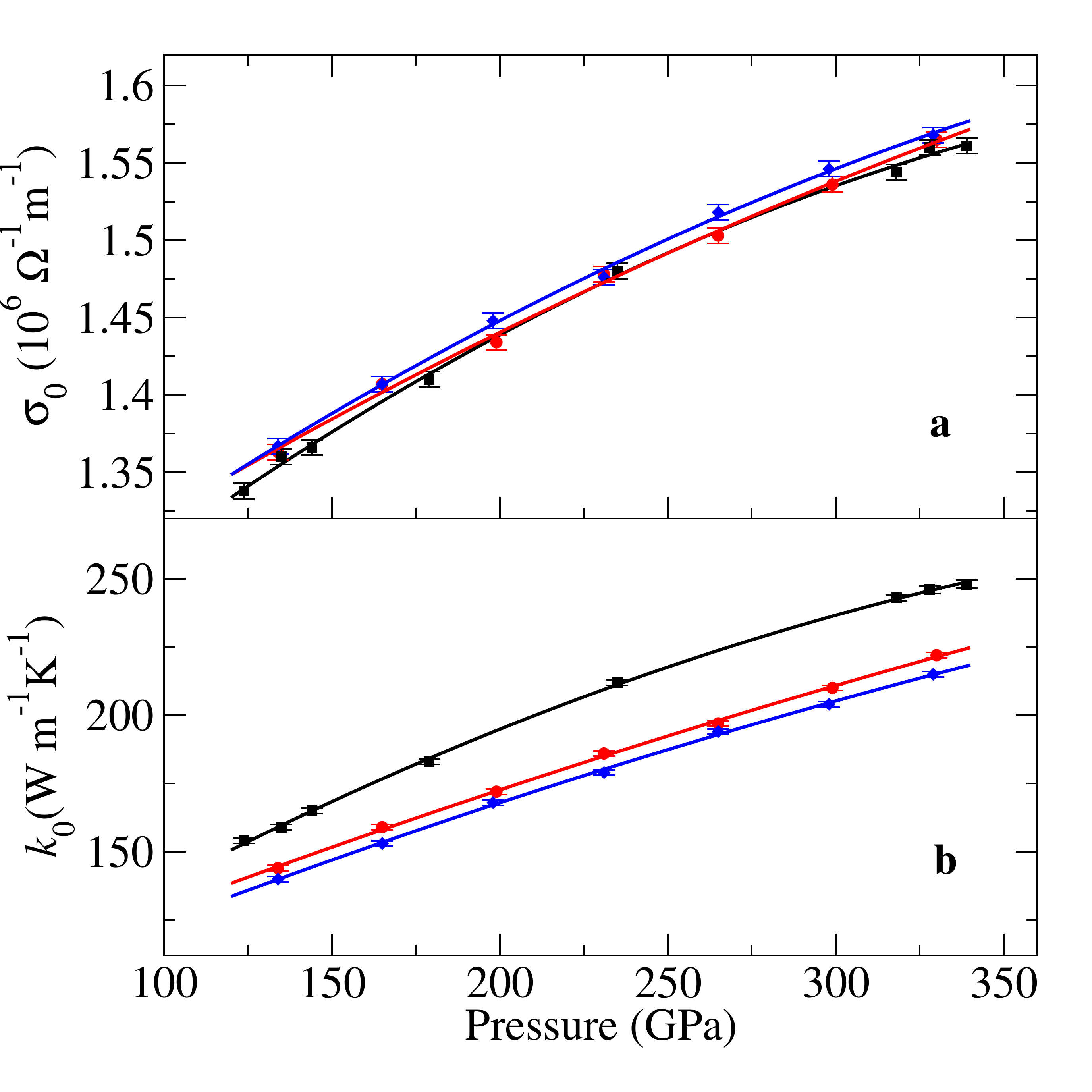}
\caption{\label{fig:cond1} Electrical (a) and thermal (b) conductivity
  of liquid iron at Earth's core conditions, computed on the FERRO
  (black), CORE5700 (red) and CORE5500 (blue) adiabats. Lines are
  quadratic fits to the first principle raw data (symbols). Error bars
  (2 s.d.) are estimated from the scattering of the data obtained from
  40 statistical independent configurations.  Results are obtained
  with cells including 157 atoms and the single {\bf k}-point
  (1/4,1/4,1/4), which are sufficient to obtain convergence within
  less than 1\%.}
\end{figure}

\begin{figure}
\includegraphics[width=4.0in]{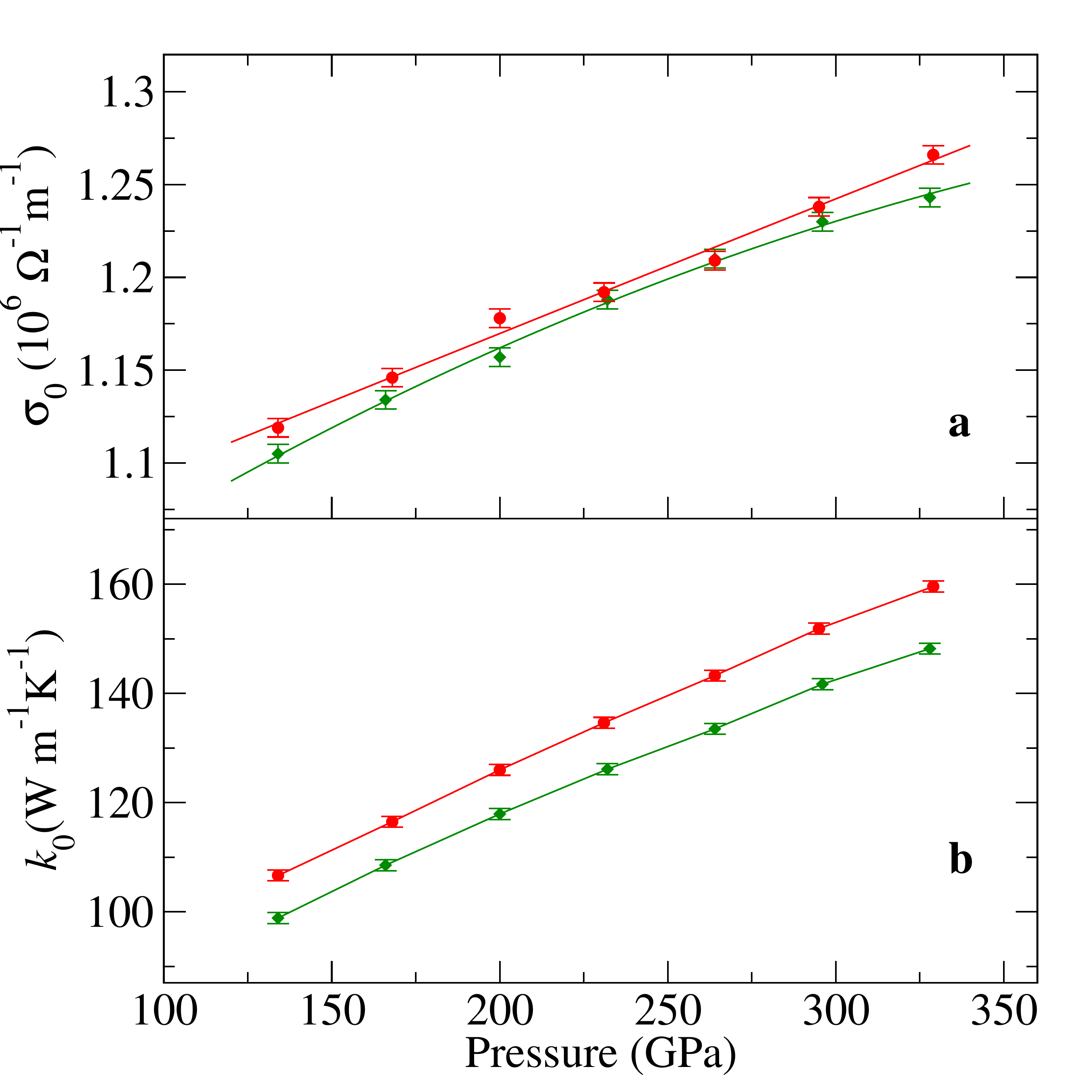}
\caption{\label{fig:condcore} Electrical (a) and thermal (b)
  conductivity of liquid Fe$_{0.82}$Si$_{0.10}$O$_{0.08}$ (red) and
  liquid Fe$_{0.79}$Si$_{0.08}$O$_{0.13}$ (green) mixtures on the CORE5700 and
  CORE5500 adiabats, respectively.  Error bars (2 s.d.) are estimated
  from the scattering of the data obtained from 40 statistical
  independent configurations.  Results are obtained with cells
  including 157 atoms and the single {\bf k}-point (1/4,1/4,1/4),
  which are sufficient to obtain convergence within less than 1\%.}
\end{figure}

The electrical and thermal conductivity of pure iron on the three
adiabats shown in Fig.~\ref{fig:cond1} have been reported
before~\cite{pozzo12}, where we also mentioned that preliminary
results showed that light impurities have the effect of reducing the
conductivities of pure iron by $\sim 30 \%$. Here in
Fig.~\ref{fig:condcore} we report the full calculations on the
CORE5500 and the CORE5700 adiabats for the mixtures. It is clear that
indeed the effect of the light impurities is that of reducing the
conductivities of pure iron by the reported amount of $\sim 30\%$, and
that the slightly different composition of the two mixtures has
roughly the same reduction effect.  The values for the electrical and
thermal conductivities are in the range 1.1-1.3 $\times 10^6
\Omega^{-1}$ m$^{-1}$ and 100-160 W m$^{-1}$ K$^{-1}$, respectively,
with the low/high values corresponding to CMB/ICB pressure-temperature
conditions.

The Lorenz parameter is roughly constant on all three adiabats,
indicating that the Wiedemann-Franz law is valid throughout the
core. For pure iron, the Lorenz parameter varies between 2.47 $\times
10^{-8}$ W $\Omega$ K$^{-2}$ and 2.51 $\times 10^-8$ W $\Omega$
K$^{-2}$, only slightly higher than its ideal value of 2.44 $\times
10^{-8}$ W $\Omega$ K$^{-2}$, while for the mixtures it is reduced in
the range 2.17-2.24 $\times 10^{-8}$ W $\Omega$ K$^{-2}$.

The values we find for the conductivities and the Lorenz number are in
broad agreement with those recently reported by de Koker et
al.~\cite{Koker12}.  Our electrical conductivities are also in fairly
good agreement with the experimental data for FeSi up to 140 GPa of
Matassov~\cite{Matassov77} and for Fe$_{0.94}$O up to 155 GPa of
Knittle et al.~\cite{Knittle86}, who reported values in the range
1.0-1.2 $\times 10^6 \Omega^{-1}$ m$^{-1}$.  Our thermal conductivities
are also in agreement with recent experimental findings of Hirose et
al.~\cite{Hirose11} who reported values for the top of the outer core
in the range 90-130 W m$^{-1}$ K$^{-1}$.

All results are summarised in Table I, II and III.

\section{Implications for the Earth}\label{sec:implications}

The estimates of thermal and electrical conductivity in
Tables~\ref{tab:2} and \ref{tab:3} are $2-3$ times higher than those
currently used in the geophysical literature \citep[e.g.][]{Stacey07a,
  Nimmo07}. The high thermal conductivity in particular has
significant implications for the evolution of the core and the dynamo
process generating the Earth's magnetic field. The convective motions
in the outer core that are responsible for the Earth's dynamo are
driven by a combination of thermal and chemical buoyancy sources.  The
strength of thermal driving is measured by the amount of excess heat
that cannot be conducted down the adiabatic gradient; higher thermal
conductivity increases adiabatic conduction and therefore decreases
the effectiveness of thermal buoyancy relative to chemical
buoyancy. Maintaining the same magnetic field with less available
thermal buoyancy requires a faster core cooling rate or a higher
concentration of radiogenic elements in the core or a combination of
the two. Moreover, a faster cooling rate implies that the inner core,
which is already thought to be a relatively young feature of the Earth
(age $\sim$ 1Ga,~\cite{Labrosse01}), is even younger.

The reduction in thermal buoyancy due to increased adiabatic
conduction is so great that parts of the liquid core are very likely
to be subcritical to thermal convection. Chemical convection, driven
by release of light elements at the boundary of the solid inner core
on freezing, may still be able to stir these regions, heat being
transported downwards to maintain the adiabatic
gradient~\cite{Loper78}.  Near the top of the core chemical convection
weakens because of the barrier of the core-mantle boundary (CMB); here
the liquid is likely to be density stratified with little or no
vertical motion at all~\cite{pozzo12}. Such a stable layer could be
detected observationally, either by seismology or by its effect on the
geomagnetic field.

It has been suggested that the heat convected away from the inner core
may vary so much from place to place that the surface may be melting
in some places, freezing in others, with a net growth from freezing
over the whole surface~\cite{Gubbins11}. The increase in thermal
conductivity with depth (Tables~\ref{tab:2} and \ref{tab:3}) increases
the heat conducted away from the ICB, making the chance of melting
rather less likely. It also reduces the vigour of thermal convection
deep in the core.



The main consequence of a higher electrical conductivity is to
lengthen the ohmic diffusion time, the main time scale used when
interpreting geomagnetic and paleomagnetic phenomena. Doubling or
trebling the time scale affects virtually all interpretations to a
certain extent: the dipole decay time increases from $25$ kyr to $75$
kyr. For example, it improves the "frozen-flux" assumption, commonly
used to interpret recent geomagnetic secular variation~\cite{Holme07},
changes on time scales of decades-to-centuries; a polarity reversal
that can take anything from 1 to 10 kyr now appears fast on the
diffusion time scale. The magnetic Reynolds number for the dynamo is
also larger, making dynamo action possible with lower flow
speeds. Lastly, the inner core has been thought to provide a
stabilising influence because its ohmic diffusion time (5,000 years)
is longer than the advective time in the outer core (500 years)~\cite{Gubbins99}. 
The higher electrical conductivity reported in this paper increases the
diffusion time considerably, making the stabilising effect even
stronger.

The ab-initio results also have implications for numerical models of
the geodynamo.  The equations governing the geodynamo process are
usually cast into nondimensional form; it is well known that
fundamental aspects of the solutions depend critically on the values
of the dimensionless parameters. Common dimensionless parameters are
the thermal Prandtl number $Pr_T=\nu/D_T$, the ratio of viscous and
thermal diffusion, the magnetic Prandtl number $Pm = \nu / D_{B}$, the
ratio of viscous and magnetic diffusion, and the chemical Prandtl
number $Pr_{\rm X} = \nu/D_{\rm X}$, the ratio of viscous diffusion
and mass diffusion for species $\rm{X}=$O,Si. Here $\nu = \eta / \rho$
is the kinematic viscosity, $D_T = \kappa_0 / \rho C_p$ is the thermal
diffusivity where $C_p$ is the specific heat at constant pressure, and
$D_B = 1/(\mu_0 \sigma_0)$ is the magnetic diffusivity where $\mu_0$
is the permeability of free space.

In Table~\ref{tab:4} we list values for the Prandtl numbers using the
results of our ab initio calculations (Tables~\ref{tab:2} and
\ref{tab:3}) at the CMB (P=134 GPa) and ICB together with errors. 
$Pm$ is slightly higher than a previous estimate of $6\times 10^{-7}$~\cite{Gubbins}, 
but this is still too low to be achievable in a simulation in the near future. Values of
$Pr_T$ are lower than the value $Pr_T=0.1$ commonly assumed for liquid
metals and can certainly be accommodated in current geodynamo
models. The values of $Pr_{\rm Si}$ and $Pr_{\rm O}$ differ
significantly and are larger than previous
estimates~\cite{Gubbins}. The Lewis number $Le = Pr_{\rm O}/Pr_T$, an
important parameter in dynamo models driven by both thermal and
compositional buoyancy, is found to be $O(10^3)$.

\section{Conclusions}\label{sec:conclusions}
In this work we have studied the transport properties for two liquid
silicon-oxygen-iron mixtures, i.e., Fe$_{0.79}$Si$_{0.08}$O$_{0.13}$
and Fe$_{0.82}$Si$_{0.10}$O$_{0.08}$, at Earth's core conditions using
DFT-KG ab-initio theoretical calculations. We find that both the
thermal and electrical conductivity of the mixtures are higher than
previous estimates, the former being in the range of 100-160 W
$m^{-1}$ K$^{-1}$ and the latter in the range of 1.1-1.3 $\times 10^6
\Omega^{-1}$ m$^{-1}$.  The validity of the Wiedemann-Franz law is
found to be satisfied quite accurately for both iron mixtures at core
conditions, the Lorenz parameter being roughly constant within the
range 2.17-2.24 $\times 10^-8$ W $\Omega$ K$^{-2}$, only slightly
higher than the ideal value of 2.44 $\times 10^-8$ W $\Omega$
K$^{-2}$.  The values we find for the conductivities are 2 to 3 times
higher than those which have been used until now and have profound
implications for our understanding of the outer core evolution and the
geodynamo. In particular, the inner core is now younger than
previously thought, the top of the core is very likely to be thermally
stably stratified, while the possibility that the inner core is
partially melting is less likely. Finally, we list values for the
dimensionless input parameters used in geodynamo simulations,
calculated directly from the ab initio calculations.

\section*{Acknowledgements}
The work of MP was supported by a NERC grant number
NE/H02462X/1. Calculations were performed on the HECToR service in the
U.K. and also on Legion at UCL as provided by research computing.
CD is supported by a Natural Environment Research Council personal 
fellowship, NE/H01571X/1.


\begin{table}
\begin{tabular}{lcccccccccc}
\hline
P & T & $\rho$& $v_b$ & $D_{Fe}$& & & $\eta$ & $\sigma_0$ & $\kappa_0$ & L\\
\hline
GPa & K & g cm$^{-3}$ & km s$^{-1}$ & 10$^{-8}$m$^2$s$^{-1}$ & & & mPa s& $ 10^6 \Omega^{-1}$ m$^{-1}$ & W m$^{-1}$ K$^{-1}$ & $ 10^{-8}$ W $\Omega$ K$^{-2}$\\
\hline
339 & 6420 & 13.16 & 10.48 &0.50(1) &  &  & 12.4(13) & 1.561(5)&248(1)&2.47(2)\\
328 & 6350 & 12.95 & 10.36 &0.50(1) & &  & 11.9(9) & 1.560(5)&246(1)&2.48(2)\\
318 & 6282 & 12.86 & 10.27 & 0.53(1) &  & & 11.4(8) & 1.544(5)&243(1)&2.50(2)\\
240 & 5700 & 12.05 & 9.45 & 0.55(1) &  & & 10.0(8) & 1.480(5)&212(1)&2.51(2)\\
178 & 5200 & 11.30 & 8.70 & 0.57(1) &  &  & 7.7(8) & 1.410(5)&183(1)&2.50(2)\\
144 & 4837 & 10.83 & 8.23 & 0.60(1) &  &  & 6.6(5) & 1.366(5)&165(1)&2.50(2)\\
135 & 4700 & 10.69 & 8.10 & 0.58(1) &  &  & 6.9(4) & 1.360(5)&159(1)&2.49(2)\\
124 & 4630 & 10.52 & 7.93 & 0.59(1) &  &  & 6.2(4) & 1.338(5)&154(1)&2.49(2)\\
\hline
\end{tabular}
\caption{ Pressure (P), temperature (T), density ($\rho$), bulk sound velocities ($v_b$), iron diffusion coefficient ($D_{Fe}$), viscosity ($\eta$), electrical conductivity ($\sigma_0$), thermal conductivity ($\kappa_0$) and Lorenz parameter (L) for pure liquid iron on the FERRO adiabat. }
\label{tab:1}
\end{table}

\begin{table}
\begin{tabular}{lcccccccccc}
\hline
P & T & $\rho$& $v_b$ & $D_{Fe}$& $D_{Si}$& $D_{O}$ & $\eta$ & $\sigma_0$ & $\kappa_0$ & L\\
\hline
GPa & K & g cm$^{-3}$ & km s$^{-1}$ & 10$^{-8}$m$^2$s$^{-1}$ & 10$^{-8}$m$^2$s$^{-1}$ & 10$^{-8}$m$^2$s$^{-1}$ & mPa s& $ 10^6 \Omega^{-1}$ m$^{-1}$ & W m$^{-1}$ K$^{-1}$ & $ 10^{-8}$ W $\Omega$ K$^{-2}$\\
\hline
329 & 5700 & 12.16 & 10.52 &0.43(1) & 0.41(2) & 0.98(6) & 11.7(10) & 1.266(5)&160(1)&2.21(2)\\
295 & 5490 & 11.83 & 10.19 &0.43(1) & 0.41(2) & 0.99(4) & 11.2(7) & 1.238(5)&152(1)&2.23(2)\\
264 & 5280 & 11.53 & 9.88 & 0.44(1) & 0.44(2) & 1.12(6) & 11.6(10) & 1.209(5)&143(1)&2.24(2)\\
231 & 5050 & 11.18 & 9.52 & 0.43(1) & 0.41(2) & 1.11(6) & 8.9(6) & 1.192(5)&135(1)&2.24(2)\\
200 & 4810 & 10.82 & 9.16 & 0.44(1) & 0.47(2) & 1.29(5) & 8.2(7) & 1.178(5)&126(1)&2.22(2)\\
168 & 4550 & 10.42 & 8.76 & 0.47(1) & 0.46(2) & 1.28(5) & 7.6(6) & 1.146(5)&117(1)&2.23(2)\\
134 & 4260 & 9.958 & 8.29 & 0.50(1) & 0.52(2) & 1.31(5) & 6.8(6) & 1.119(5)&107(1)&2.24(2)\\
\hline
\end{tabular}
\caption{ Pressure (P), temperature (T), density ($\rho$), bulk sound velocities ($v_b$), diffusion coefficients for iron ($D_{Fe}$), silicon ($D_{Si}$) and oxygen ($D_{O}$), viscosity ($\eta$), electrical conductivity ($\sigma_0$), thermal conductivity ($\kappa_0$) and Lorenz parameter (L) for the Fe$_{0.82}$Si$_{0.10}$O$_{0.08}$ liquid mixture on the CORE5700 adiabat. }
\label{tab:2}
\end{table}

\begin{table}
\begin{tabular}{lcccccccccc}
\hline
P & T & $\rho$& $v_b$ & $D_{Fe}$& $D_{Si}$& $D_{O}$ & $\eta$ & $\sigma_0$ & $\kappa_0$ & L\\
\hline
GPa & K & g cm$^{-3}$ & km s$^{-1}$ & 10$^{-8}$m$^2$s$^{-1}$ & 10$^{-8}$m$^2$s$^{-1}$ & 10$^{-8}$m$^2$s$^{-1}$ & mPa s& $ 10^6 \Omega^{-1}$ m$^{-1}$ & W m$^{-1}$ K$^{-1}$ & $ 10^{-8}$ W $\Omega$ K$^{-2}$\\
\hline
328 & 5500 & 12.04 & 10.53 &0.38(1) & 0.38(1) & 0.92(3) & 13.1(9) & 1.243(5)&148(1)&2.17(2)\\
296 & 5300 & 11.75 & 10.24& 0.37(1) & 0.38(2) & 0.90(4) & 11.4(9) & 1.230(5)&142(1)&2.17(2)\\
264 & 5095 & 11.42 & 9.93 &0.38(1) & 0.38(2) & 0.98(3) & 10.4(8) & 1.210(5)&134(1)&2.17(2)\\
232 & 4870 & 11.10 & 9.60&0.41(1) & 0.41(3) & 1.00(3) & 9.4(8) & 1.188(5)&126(1)&2.18(2)\\
200 & 4640 & 10.73 & 9.25&0.41(1) & 0.40(3) & 1.19(3) & 8.4(7) & 1.157(5)&118(1)&2.20(2)\\
166 & 4385 & 10.32 & 8.84&0.46(1) & 0.45(2) & 1.19(4) & 7.6(6) & 1.134(5)&109(1)&2.18(2)\\
134 & 4112 & 9.887 & 8.43& 0.45(1) & 0.46(2) & 1.30(7) & 6.7(8) & 1.105(5)&99(1)&2.18(2)\\
\hline
\end{tabular}
\caption{ Pressure (P), temperature (T), density ($\rho$), bulk sound velocities ($v_b$), diffusion coefficients for iron ($D_{Fe}$), silicon ($D_{Si}$) and oxygen ($D_{O}$), viscosity ($\eta$), electrical conductivity ($\sigma_0$), thermal conductivity ($\kappa_0$) and Lorenz parameter (L) for the Fe$_{0.79}$Si$_{0.08}$O$_{0.13}$ liquid mixture on the CORE5500 adiabat. }
\label{tab:3}
\end{table}

\begin{table}
\begin{tabular}{lccccc}
\hline
P &  $Pr_T$        & $Pm$ $\times 10^{-6}$ & $Pr_{\rm Si}$ & $Pr_{\rm O}$  &    $D_T$ $\times 10^{-5}$ \\ 
\hline
328 &  0.063 (0.004) & 1.7 (0.12)           & 286 (27)  & 118 (12)  &    1.7 (0.03)    \\
134 &  0.048 (0.005) & 0.9 (0.11)           & 147 (24)  & 52 (9)    &    1.4 (0.02)    \\  
\hline
329 &  0.052 (0.004) & 1.5 (0.13)           & 235 (31)  & 98 (14)   &    1.8 (0.03)    \\
134 &  0.046 (0.005) & 1.0 (0.09)           & 131 (16)  & 52 (7)    &    1.5 (0.02)    \\
\hline
\end{tabular}
\caption{Values of the thermal Prandtl number ($Pr_T$), magnetic Prandtl number ($Pm$), and 
  chemical Prandtl numbers for Silicon ($Pr_{\rm Si}$) and Oxygen ($Pr_{\rm O}$). Also reported is the thermal diffusivity $D_T$. Top section is 
  for the inner-core density jump of Masters and Gubbins~\cite{masters03}; bottom section 
  is for the PREM density jump~\cite{dziewonski81}. Error estimates are in brackets. }
\label{tab:4}
\end{table}

\end{document}